\documentclass{aastex63}

\shortauthors{Lowry et al.}

\usepackage{amsmath,bm}

\graphicspath{{./}{figures/}}

\usepackage{tikz}
\usetikzlibrary{arrows,shapes,positioning,shadows,trees}

\tikzset{
  basic/.style  = {draw, text width=4cm, drop shadow, font=\sffamily, rectangle},
  root/.style   = {basic, rounded corners=2pt, thin, align=center, fill=blue!10},
  level 2/.style = {basic, rounded corners=6pt, thin,align=center, fill=pink!40, text width=11em},
  level 3/.style = {basic, thin, align=center, fill=white!60, text width=8em}
}

\usepackage[utf8]{inputenc}
\usepackage{dirtytalk}

\begin{document}

\title{Clarissa Family Age from the Yarkovsky Effect Chronology}

\correspondingauthor{Vanessa Lowry}
\email{vanessa\_lowry@knights.ucf.edu}

\author{Vanessa C. Lowry}
\affiliation{University of Central Florida, 4000 Central Florida Blvd, Orlando, FL 32816, USA}

\author{David Vokrouhlick\'y}
\affiliation{Astronomical Institute, Charles University, Prague, V Hole\v{s}ovi\v{c}k\'ach 2, CZ 18000, Prague 8, Czech Republic}

\author{David Nesvorn\'y}
\affiliation{Department of Space Studies, Southwest Research Institute, 1050 Walnut St., Suite 300, Boulder, CO 80302, USA}

\author{Humberto Campins}
\affiliation{University of Central Florida, 4000 Central Florida Blvd, Orlando, FL 32816, USA}

\begin{abstract}

The Clarissa family is a small collisional family composed of primitive C-type asteroids. It is located in a dynamically 
stable zone of the inner asteroid belt. In this work we determine the formation age of the Clarissa family by modeling planetary 
perturbations as well as thermal drift of family members due to the Yarkovsky effect. Simulations were carried out using 
the SWIFT-RMVS4 integrator modified to account for the Yarkovsky and Yarkovsky–O'Keefe–Radzievskii–Paddack (YORP) effects. We ran multiple simulations starting 
with different ejection velocity fields of fragments, varying proportion of initially retrograde spins, and also tested different 
Yarkovsky/YORP models. Our goal was to match the observed orbital structure of the Clarissa family which is notably asymmetrical
in the proper semimajor axis, $a_{\rm p}$. The best fits were obtained with the initial ejection velocities $\lesssim$ 20 m s$^{-1}$ of 
diameter $D \simeq 2$ km fragments, $\sim$ 4:1 preference for spin-up by YORP, and assuming that $\simeq$ 80\% of small 
family members initially had retrograde rotation. The age of the Clarissa family was found to be $t_{\rm age} = 56 \pm 6$ Myr for 
the assumed asteroid density $\rho = 1.5$ g cm$^{-3}$.  Small variation of density to smaller
or larger value would lead to 
slightly younger or older age estimates. This is the first case where the Yarkovsky effect chronology has been successfully applied 
to an asteroid family younger than 100 Myr.

\end{abstract}


\section{Introduction} \label{sec:intro}

Asteroid families consist of fragments produced by catastrophic and cratering impacts on parent bodies (see \citeauthor{2015aste.book..297N} \citeyear{2015aste.book..509V} for 
a review). The fragments produced in a single collision, known as family members, share similar proper semimajor axes ($a_{\rm p}$), 
proper eccentricities ($e_{\rm p}$), and proper inclinations ($i_{\rm p}$) \citep{2002aste.book..603K, 2016MNRAS.457.1332C}.  
Family members are also expected to have spectra that indicate similar mineralogical composition to the parent body \citep{2015aste.book..323M}.  After their formation, families experience collisional evolution \citep{Marzari1995}, which may 
cause them to blend into the main belt background, and evolve dynamically \citep{Bottke2001}.  Since the collisional lifetime of a 2 km sized body in the main belt is greater than 500 Myr \citep{2005Icar..179...63B}, which is nearly 10 times longer than the Clarissa family's age (Section~\ref{sec:results}), we do not need to account for collisional evolution. Instead, we only consider dynamical evolution of the Clarissa family to explain its current orbital structure and constrain its formation conditions and age.

\begin{figure}[ht!]
\epsscale{0.8}
\plotone{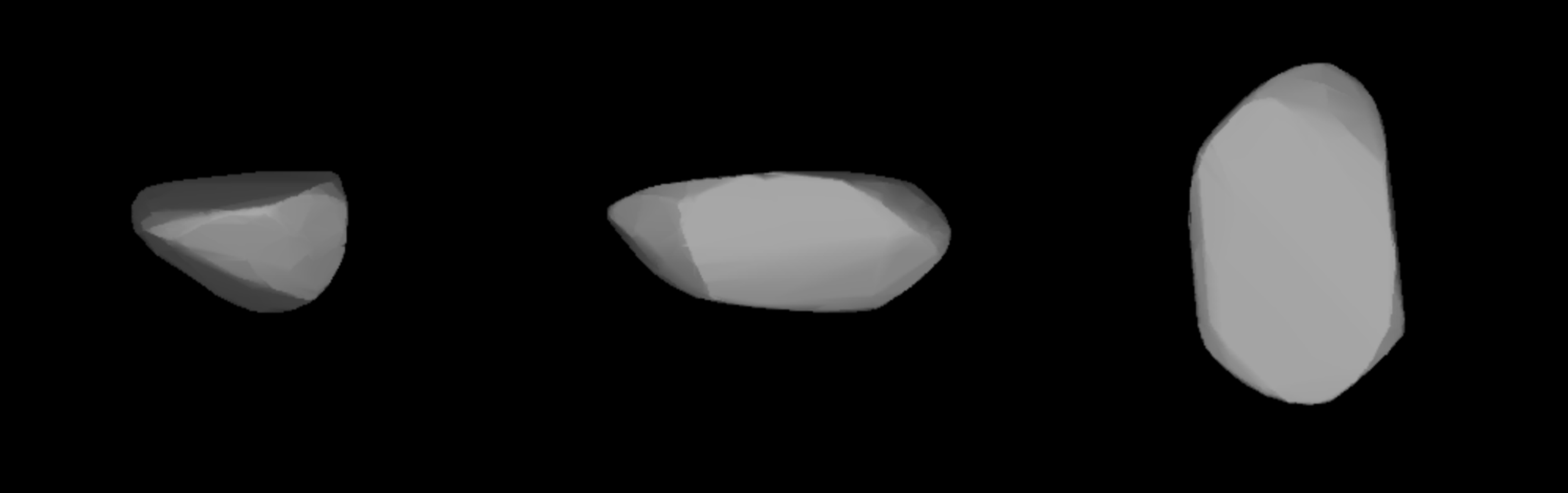}
\caption{Three-dimensional shape of (302) Clarissa from light-curve inversion (\citeauthor{2010A&A...513A..46D} \citeyear{2010A&A...513A..46D}, \citeyear{2011Icar..214..652D}; \citeauthor{2011A&A...530A.134H} \citeyear{2011A&A...530A.134H}; https://astro.troja.mff.cuni.cz/projects/damit/). Equatorial views along the principal 
axes of inertia are shown in the left and middle panels.  The polar view along the shortest inertia axis is shown on the right (view from the direction of the rotation pole).} \label{fig:fig1}
\end{figure}

The Clarissa family is one of small, primitive (C or B type in asteroid taxonomy) families in the inner asteroid belt \citep{2018A&A...610A..25M}.
Of these families, Clarissa has the smallest extent in semimajor axis suggesting that it might be the youngest. This  
makes it an interesting target of dynamical study.  Before discussing the Clarissa family in depth, we summarize what is known about 
its largest member, asteroid (302) Clarissa shown in Figure \ref{fig:fig1}.  Analysis of dense and sparse photometric data shows that (302) Clarissa has a retrograde spin 
with rotation period $P=14.4797$ hr, pole ecliptic latitude $-72^{\circ}$, and two possible solutions for pole ecliptic longitude, $28^{\circ}$ 
or $190^{\circ}$ \citep{2011A&A...530A.134H}.  The latter solution was excluded once information from stellar occultations became available
\citep{2010A&A...513A..46D, 2011Icar..214..652D}. The stellar occultation also provided a constraint on the volume-equivalent diameter $D=43 \pm 4$ km of
(302) Clarissa \citep{2010A&A...513A..46D, 2011Icar..214..652D} improving on an earlier estimate of $39 \pm 3$~km from the analysis of IRAS data \citep{2004PDSS...12.....T}.  

\begin{figure}[ht!]
\epsscale{0.75}
\plotone{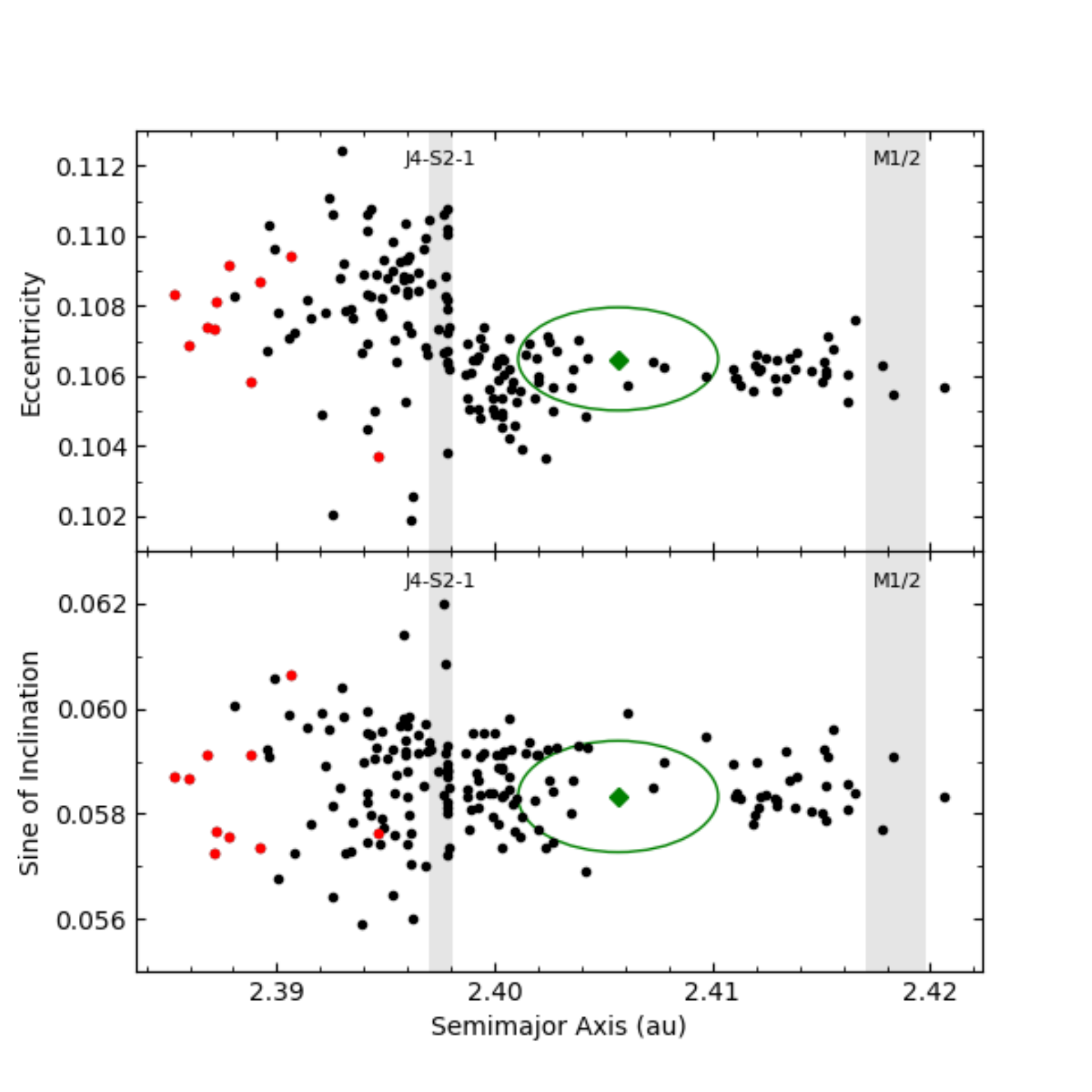}
\caption{Orbital structure of the Clarissa family: $e_{\rm p}$ vs. $a_{\rm p}$ (top panel), sine of $i_{\rm p}$ vs. $a_{\rm p}$ (bottom panel). (302) Clarissa is highlighted 
by the green diamond.  
The locations of principal mean motion resonances are indicated by the gray shaded regions with the three-body J4-S2-1 resonance on the left and the exterior M1/2 resonance
with Mars on the right. The resonance widths were computed using software available from \cite{2014Icar..231..273G} (see also \citeauthor{NesvornyMorbi1998} \citeyear{NesvornyMorbi1998}).  It can be noted that the dispersion of the fragments surrounding (302) Clarissa is narrow in $e_{\rm p}$, but sunward of the 
J4-S2-1 resonance, fragments are more dispersed in $e_{\rm p}$.  The red symbols mark potential interlopers in the family based on their location in the ($a_{\rm p}, H$) plane (see Figure \ref{fig:fig3}).  The green 
ellipses indicate a region in proper element space where the initial fragments would land assuming: (i) isotropic ejection from the parent body, (ii) $f \simeq \omega \simeq 90^\circ$ (see Appendix \ref{Ap:A}), and (iii) velocity of 20 m s$^{-1}$. This is equal to the escape velocity from (302) Clarissa ($v_{\rm esc} = 19.7 \pm 1.8$ m s$^{-1}$ for $D = 43 \pm 4$ km and 
bulk density $\rho = 1.5$ g cm$^{-3}$). \label{fig:fig2}}
\end{figure}

The 179 family members of the Clarissa family are tightly clustered around (302) Clarissa \citep{2015aste.book..297N}.  The synthetic proper elements of the Clarissa family members were obtained from the Planetary Data System (PDS) node \citep{2015PDSS..234.....N}.\footnote{Re-running the hierarchical clustering method  (\citeauthor{1990AJ....100.2030Z} \citeyear{1990AJ....100.2030Z, 1994AJ....107..772Z}) on the most recent proper element catalog results in only a slightly larger membership of the Clarissa family. As the orbital structure of the family remains the same, we opt to use the original PDS identification.}
Figure~\ref{fig:fig2} shows the projection of the Clarissa family onto the $(a_{\rm p}, e_{\rm p})$ and $(a_{\rm p}, \sin{i_{\rm p}})$ planes. For comparison, we also indicate in Fig. \ref{fig:fig2} the initial distribution of fragments if they were ejected at speeds equal to the escape speed from (302) Clarissa. 

To generate these initial distributions we adopted the argument of perihelion $\omega \simeq 90^{\circ}$ and true anomaly $f \simeq 90^{\circ}$, both given at the moment of the parent body breakup \citep{1984Icar...59..261Z, 2006IAUS..229..289N, 2006Icar..182..118V} (see Appendix \ref{Ap:A}). 
Other choices of these (free) parameters would lead to ellipses in Figure \ref{fig:fig2} that would be tilted in the $(a_{\rm p}, e_{\rm p})$ projection 
and/or vertically squashed in the $(a_{\rm p}, i_{\rm p})$ projection \citep{2017AJ....153..172V}.  Interestingly, the areas delimited by the green 
ellipses in Fig.~\ref{fig:fig2} contain only a few known Clarissa family members. We interpret this as a consequence of the dynamical spreading of the Clarissa family by the Yarkovsky effect. 

\begin{figure}[ht!]
\epsscale{0.6}
\plotone{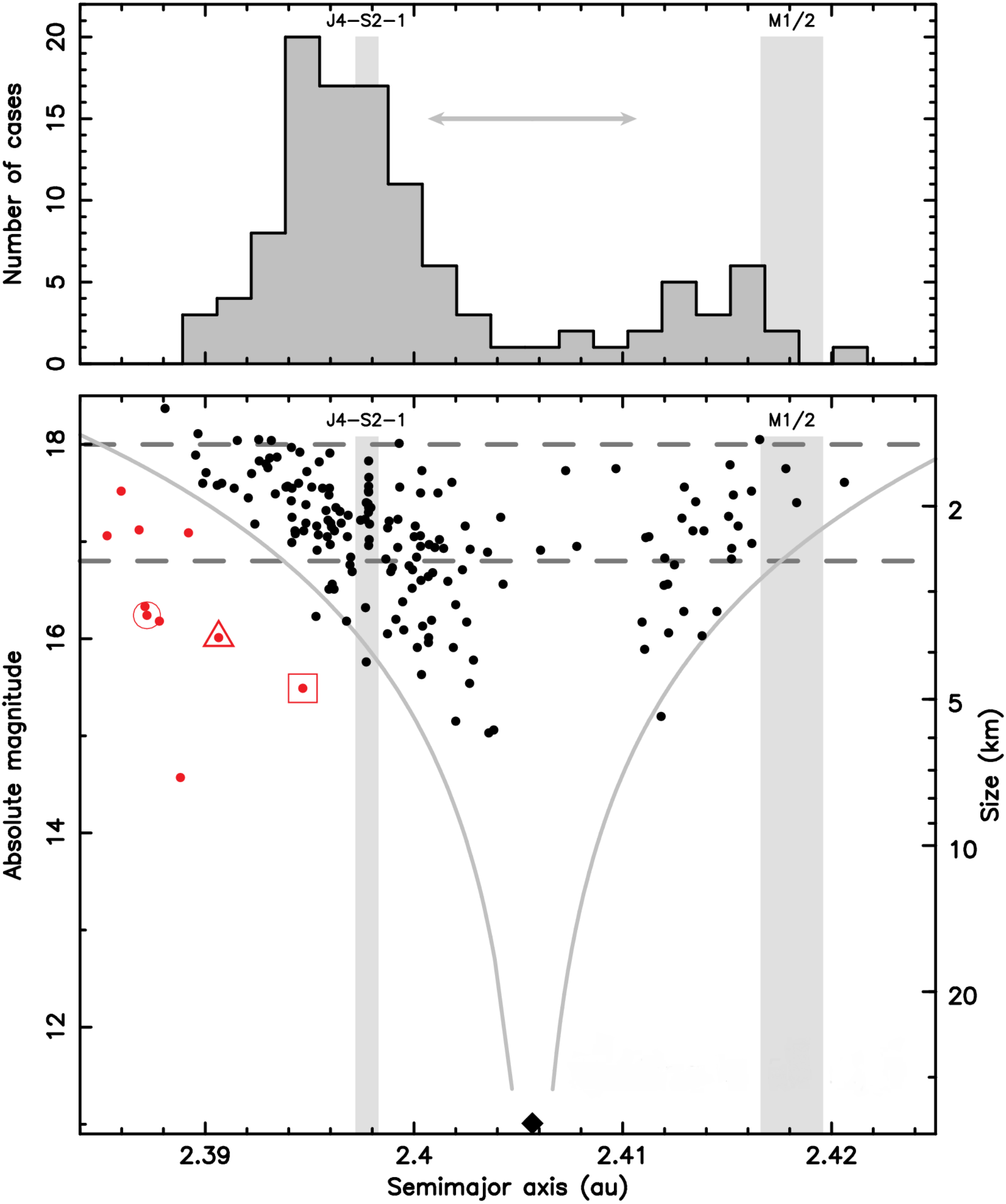}
\caption{Clarissa family distribution in $a_{\rm p}$ and $H$ (bottom panel). (302) Clarissa is highlighted by the black diamond ($D = 43 \pm 4$ km; \citeauthor{2011Icar..214..652D} \citeyear{2011Icar..214..652D}).  Small family members ($H>$ 17) are missing in the center of the family and they are pushed toward the borders. The curves are calculated by Eq. (\ref{eq:H}), where $a_{\rm c}=2.4057$~au corresponds to (302) Clarissa and $C = (5 \pm 1) \times 10^{-6}$~au \citep{2015aste.book..297N}.  Asteroids located far outside this envelope, shown in red, are suspected interlopers. The top panel shows the $a_{\rm p}$ distribution of family members (suspected interlopers excluded) with $16.8<H<18$ (corresponding to $D\simeq2$ km).
The gray arrow indicates a range of $a_{\rm p}$ values that would be initially populated by Clarissa family members (assumed ejection speeds $\lesssim 20$ m s$^{-1}$).} 
\label{fig:fig3}
\end{figure}

Immediately following the impact on (302) Clarissa, the initial spread of fragments reflects their ejection velocities.    
We assume that the Clarissa family was initially much more compact than it is now (e.g., the green ellipses in Fig.~\ref{fig:fig2}). As the family members drifted by the 
Yarkovsky effect, the overall size of the family in $a_{\rm p}$ slowly expanded. It is apparent that the Clarissa family has undergone Yarkovsky drift since
there is a depletion of asteroids in the central region of the family in Figure \ref{fig:fig3}.

There are no major resonances in the immediate orbital neighborhood of (302) Clarissa. The $e_{\rm p}$ and $i_{\rm p}$ values of family members therefore remained initially unchanged. Eventually, the family members reached the principal mean motion resonances,
most notably the J4-S2-1 three-body resonance at $\simeq 2.398$~au \citep{NesvornyMorbi1998} which can change  $e_{\rm p}$ and $i_{\rm p}$. This presumably 
contributed to the 
present orbital structure of the Clarissa family, where members with $a_{\rm p}<2.398$ au have significantly larger spread in $e_{\rm p}$ and $i_{\rm p}$ than those
with $a_{\rm p}>2.398$ au. Note, in addition, that there are many more family members sunward from (302) Clarissa, relative to those on the other side 
(Fig. \ref{fig:fig3}). We discuss this issue in more detail in Section \ref{sec:results}.

Figure \ref{fig:fig3} shows the absolute magnitudes $H$ of family members as a function of $a_{\rm p}$.  We use the mean WISE albedo of the Clarissa family, 
$p_{\rm V}=0.056 \pm 0.017$ \citep{2013ApJ...770....7M, 2015aste.book..323M} to convert $H$ to $D$ (shown on the right ordinate). As often seen in asteroid families, 
the small members of the Clarissa family are more dispersed in $a_{\rm p}$ than the large members. The envelope of the distribution in $(a_{\rm p},H)$ is 
consequently ``V'' shaped \citep{2006Icar..182..118V}. The small family members also concentrate toward the extreme $a_{\rm p}$ values,  while there is a lack of
asteroids in the family center, giving the family the appearance of ears. This feature has been shown to be a consequence of the YORP effect, which produces a 
perpendicular orientation of spin axis relative to the orbital plane and maximizes the Yarkovsky drift (e.g., \citeauthor{2006Icar..182..118V} \citeyear{2006Icar..182..118V}). Notably, the observed 
spread of small family members exceeds, by at least a factor of two, the spread due to the escape velocity from (302) Clarissa, and the left side of the family in Fig. 
\ref{fig:fig3} is overpopulated by a factor of $\sim$4.  

The solid gray curves in the bottom panel of Figure \ref{fig:fig3} delimit the boundaries wherein most family members are located.  The curves in the figure are 
calculated from the equation  
\begin{equation} 
H = 5\log_{10}\left(|a_{\rm p}-a_{\rm c}|/C\right),
\label{eq:H}
\end{equation}
where $a_{\rm c}$ is the family center and $C$ is a constant. The best fit to the envelope of the family is obtained with $C = (5 \pm 1) \times 10^{-6}$ au.
As explained in  \cite{2015aste.book..297N}, the constant $C$ is an expression of (i) the ejection velocity field with velocities inversely proportional 
to the size, and (ii) the maximum Yarkovsky drift of fragments over the family age. It is difficult to decouple these two effects without detailed 
modeling of the overall family structure (Sections \ref{sec:Analysis} and \ref{sec:results}). Ignoring (i), we can crudely estimate the Clarissa family age. For that we use
\begin{equation} \label{eq:eqt}
t_{\rm age} \simeq 1\, {\rm Gy} 
\left(\frac{C}{10^{-4} \ {\rm au}}\right)
\left(\frac{a}{2.5\ {\rm au}}  \right)^2 
\left(\frac{\rho}{2.5\ {\rm g} \, {\rm cm}^{-3}}\right) 
\left(\frac{0.2}{p_V}  \right)^{1/2}
\end{equation}
from \cite{2015aste.book..297N}, where $\rho$ is the asteroid bulk density and $p_{\rm V}$ is the visual albedo.  For the Clarissa family we 
adopt $C= (5 \pm 1)$ $\times 10^{-6}$ au, and values typical for a C-type asteroid: $\rho = 1.5$ g cm$^{-3}$,  $p_V = 0.05$,  and find $t_{\rm age} \simeq 56 \pm 11$ Myr.
Using similar arguments \cite{2019MNRAS.484.1815P} estimated that the Clarissa family is $\sim$ 50-80 My old. Furthermore, \cite{2015Icar..247..191B} 
suggested $t_{\rm age} \simeq$ 60 Myr for the Clarissa family, but did not attach an error bar to this value.

Objects residing far outside of the curves given by Eq. (\ref{eq:H}) are considered interlopers (marked red in Figs. \ref{fig:fig2} and \ref{fig:fig3}) and are not included 
in the top panel of Fig. \ref{fig:fig3}.  Further affirmation that these objects are interlopers could be obtained from spectroscopic studies
(e.g., demonstrating that they do not match the family type; \citeauthor{2006Icar..182...92V} \citeyear{2006Icar..182...92V}). The spectroscopic data for the Clarissa family are sparse, however, 
and do not allow us to confirm interlopers. We mention asteroid (183911) 2004 CB100 (indicated by a red triangle) which was found to be a spectral type X (likely an interloper).  Asteroid (36286) 2000 EL14 (indicated by a red square in Fig. \ref{fig:fig3}) has a low albedo $p_{\rm V} \simeq 0.06$ in the WISE 
catalog \citep{2011ApJ...741...68M} and \cite{2018A&A...610A..25M} found it to be a spectral type C.  Similarly, asteroid (112414) 2000 NV42 (indicated by a red 
circle) was found to be a spectral type C.  These bodies may be background objects although the background of primitive bodies in the inner main belt is not large.  The absolute magnitudes $H$ could be determined particularly badly (according to \cite{2012Icar..221..365P} the determination of $H$ may have an uncertainty accumulated up to a magnitude), but it is not clear why only these two bodies of spectral type C would have this bias.

\begin{figure}[ht!]
\epsscale{0.8}
\plotone{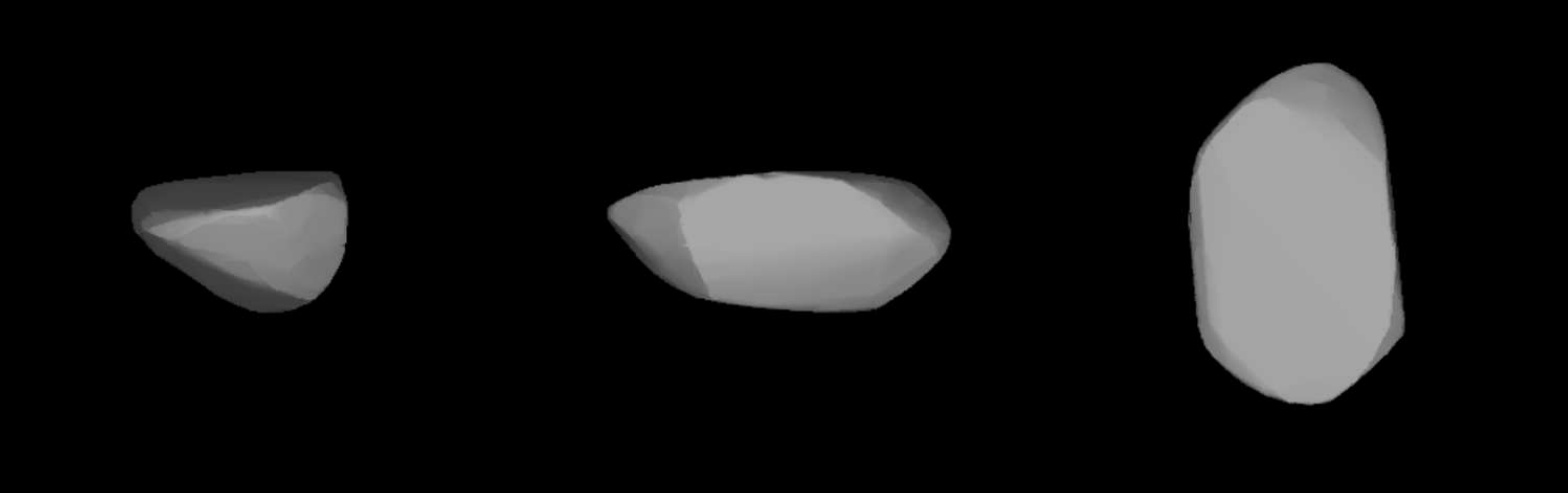}
\caption{Cumulative absolute magnitude $H$ distribution of 169 Clarissa family members (10 suspected interlopers removed; Fig. \ref{fig:fig3}). The large dot indicates (302) Clarissa. The gray reference line is $N(<\!\!H) \propto 10^{\gamma H}$ with $\gamma = 0.75$. The diameters 
are labeled on the top.} \label{fig:fig4}
\end{figure}

The absolute magnitude $H$ distribution of Clarissa family members can be approximated by a power law, $N(<\!\!H)\propto 10^{\gamma H}$ \citep{2006Icar..182...92V},
with $\gamma = 0.75$ (Fig. \ref{fig:fig4}). The relatively large value of $\gamma$ and large size of (302) Clarissa relative to other family members are indicative of a cratering event on (302) Clarissa \citep{2006Icar..182..118V}. The significant flattening of (302) Clarissa in the northern
hemisphere (Fig. \ref{fig:fig1}) may be related to the family-forming event (e.g., compaction after a giant cratering event).  This is only speculation and we should caution the reader about the uncertainties in shape modeling. In the next section, we describe 
numerical simulations that were used to explain the present orbital structure of the Clarissa family and determine its formation conditions and age.  
We take particular care to demonstrate the strength of the J4-S2-1 resonance and its effect on the family structure inside of 2.398 au.  

\section{Numerical Model}  \label{sec:numsim}

\begin{figure}[ht!]
\epsscale{0.6}
\plotone{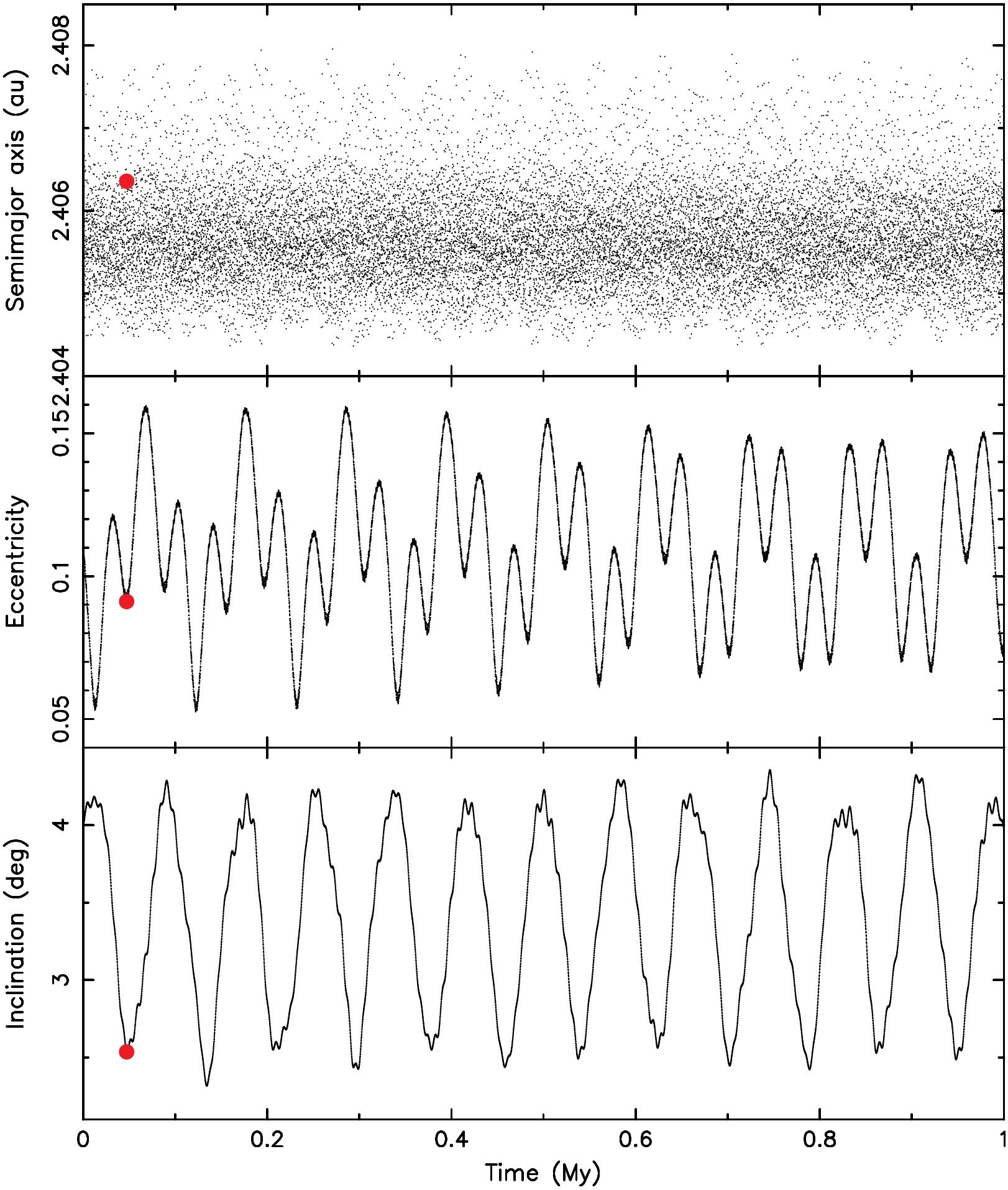}
\caption{Osculating elements of (302) Clarissa over 1 Myr into the future.  At time $t \simeq 47.35$ ky (red dot), $\omega=90^{\circ}$ and $f=90^{\circ}$.} \label{fig:fig5}
\end{figure}

As a first step toward setting up the initial conditions, we integrated the orbit of (302) Clarissa over $1$ Myr.  We determined the moment in time when 
the argument of perihelion of (302) Clarissa reached $90^{\circ}$ (note that the argument is currently near $54^{\circ}$).  Near that epoch we followed the asteroid along its orbit 
until the true anomaly reached $90^{\circ}$ (Figure \ref{fig:fig5}). This was done to have an orbital configuration compatible with $\omega \simeq f \simeq 90^{\circ}$ (see Appendix \ref{Ap:A} for more comments) that we used 
to plot the ellipses in Fig. \ref{fig:fig2}. At that epoch we recorded  (302) Clarissa's heliocentric position vector $\bm{R}$ and velocity $\bm{V}$, and added a small 
velocity change $\delta \bm{V}$ to the latter. This represents the initial ejection speed of individual members.  From that we determined the orbital elements 
using the Gauss equations with $f=90^\circ$ and $\omega=90^\circ$ \citep{1996Icar..124..156Z}.

We generated three distributions with $\delta \bm{V} = 10$, 20, and 30 m s$^{-1}$ to probe the dependence of our results on this parameter (ejection directions
were selected isotropically). Note that $\delta \bm{V}$ = 20 m s$^{-1}$ best matches the escape speed from (302) Clarissa. The assumption of a constant ejection speed of simulated 
fragments is not a significant approximation, because we restrict our modeling to $D \simeq 2$ km which is the characteristic size of most known family members.

We used the SWIFT-RMVS4 code of the SWIFT-family software written by \cite{LevSWIFT}. The code numerically integrates orbits of $N$ massive (the sun and planets) and $n$ massless 
bodies (asteroids in our synthetic Clarissa family).  For all of our simulations we included eight planets (Mercury to Neptune), thus $N=9$, and $n=500$ test family members.  
We used a time step of two days and simulated all orbits over the time span of 150 Myr. This is comfortably longer than any of the age estimates discussed in Section \ref{sec:intro}.
The integrator eliminates test bodies that get too far from the sun or impact a massive body.  Since (302) Clarissa is located in a dynamically stable zone and the 
simulation is not too long, we did not see  many particles being eliminated. Only a few particles leaked from the J4-S2-1 resonance onto planet-crossing orbits.
The Clarissa family should thus {\it not} be a significant source of near-Earth asteroids.

The integrator mentioned above takes into account gravitational forces among all massive bodies and their effect on the massless bodies.  Planetary perturbations of Clarissa 
family members include the J4-S2-1 and M1/2 resonances shown in Fig. \ref{fig:fig3} and the secular perturbations which cause oscillations of osculating 
orbital elements (Fig. \ref{fig:fig5}).  The principal driver of the long-term family evolution, however, is the Yarkovsky effect, which causes semimajor axis drift
of family members to larger or smaller values.  Thus, we extended the original version of the SWIFT code to allow us to model these thermal accelerations.  Details 
of the code extension can be found in \cite{2017AJ....153..172V} and \cite{2015Icar..247..191B} (also see Appendix \ref{Ap:B}). Here we restrict ourselves to describing the main features and 
parameters relevant to this work.

The linear heat diffusion model for spherical bodies is used to determine the thermal accelerations \citep{1999A&A...344..362V}.  We only account for the diurnal 
component since the seasonal component is smaller and its long-term effect on semimajor axis vanishes when the obliquity is $0^{\circ}$ or $180^{\circ}$.  
We use $D = 2$ km and assume a bulk density of 1.5~g~cm$^{-3}$ which should be appropriate for primitive C/B-type asteroids \citep{2015aste.book..297N, 2015aste.book..745S}.  
Considering the spectral class and size, we set the surface thermal inertia equal to 250 in the SI units \citep{2015aste.book..107D}. To model thermal drift in the 
semimajor axis we also need to know the rotation state of the asteroids: the rotation period $P$ and orientation of the spin vector $\textbf{s}$. The current rotational
states of the Clarissa family members, except for (302) Clarissa itself (see Section \ref{sec:intro}), are unknown. This introduces another degree of freedom into our model,
because we must adopt some initial distribution for both of these parameters. 
For the rotation period $P$ we assume a Maxwellian distribution between 3 and 20 hr with a peak at 6 hr based on Eq. (4) from \cite{2002aste.book..113P}. The orientation of the spin vectors was initially set to be isotropic but, as we will show, this choice turned out to be a principal obstacle in matching the orbital structure of the Clarissa family (e.g., the excess of members sunward from 
(302) Clarissa).  We therefore performed several additional simulations with non-isotropic distributions to test different initial proportions of 
prograde and retrograde spins.

The final component of the SWIFT extension is modeling the evolution of the asteroids' rotation state. For this we implement an efficient symplectic integrator  
described in \cite{2005AJ....130.1267B}. We introduce $\Delta$ which is the dynamical elipticity of an asteroid. It is an important parameter since the SWIFT code includes effects of solar gravitational torque.  We assume that $\Delta = (C - 0.5(A+B))/C,$ where $(A, B, C)$ are the principal moments of the inertia tensor; has a Gaussian 
distribution with mean 0.25 and standard deviation 0.05.  These values are representative of a population of small asteroids for which the shape models were 
obtained (e.g., \citeauthor{2002Icar..159..449V} \citeyear{2002Icar..159..449V}).  

The YORP effect produces a long-term evolution of the rotation period and direction of the spin vector \citep{2006AREPS..34..157B, 2015aste.book..509V}.  To account for that we implemented 
the model of \cite{2004Icar..172..526C} where the YORP effect was evaluated for bodies with various computer-generated shapes (random Gaussian spheroids).  
For a 2~km sized Clarissa family member, this model predicts that YORP should double  the rotational frequency over a mean timescale of $\simeq$ 80 Myr.

We define one YORP cycle as the evolution from a generic initial rotation state to the asymptotic state with very fast or very slow rotation, and 
obliquity near $0^{\circ}$ or $180^{\circ}$.  Given that the previous Clarissa family age estimates are slightly shorter than the YORP timescale quoted above,  
we expect that $D \simeq 2$ km members have experienced less than one YORP cycle.  This is fortunate because previous studies showed that modeling
multiple YORP cycles can be problematic \citep{2015Icar..247..191B, 2015A&A...579A..14V}.  As for the preference of YORP to accelerate or decelerate spins, 
\cite{2012ApJ...752L..11G} studied YORP with lateral heat conduction and found that YORP more often tends to accelerate rotation than to slow down rotation \citep[see also][for effects on obliquity]{2019AJ....157..105G}. The proportion of slow-down to spin-up cases is unknown and, for sake of simplicity, we do not model these effects in detail. Instead, we take an empirical and approximate approach. Here we investigate cases where (i) 50\% of spins accelerate and 50\% decelerate (the YORP1 model), 
and (ii) 80\% of spins accelerate and 20\% decelerate (YORP2). See Table \ref{fig:fig6} for a summary of model assumptions both physical and dynamical.   \\

\begin{table}[ht!]
\begin{center}
\begin{tabular}{ |p{9cm}||p{6cm}|  }
 \hline
 \multicolumn{2}{|c|}{\large \textbf{Summary of Physical and Dynamical Model Assumptions}} \\
 \hline
 \it \normalsize \textbf{Asteroid Physical Properties (C-type)} & \it \normalsize \textbf{Value}\\
 \hline
 (302) Clarissa's diameter & $43 \pm 4$ km  \\ \hline
 Visual albedo & $0.056 \pm 0.017 $\\
 \hline
 Bulk density   & $ 1.5$ g cm$^{-3}$ \\
 \hline
 Thermal inertia & $ 250$ J m$^{-2}$ s$^{-0.5}$ K$^{-1}$ \\ \hline
 Constant \textit{C} (see \citeauthor{2015aste.book..297N} \citeyear{2015aste.book..297N}) & $(5 \pm 1) \times 10^{-6}$ au  \\
 \hline
  & \\
 \hline 
  \it \normalsize \textbf{Dynamical Properties} & \it \normalsize \textbf{Value}  \\
  \hline
  Initial velocity field & Isotropic with 10-30 m s$^{-1}$ \\ \hline
  Initial percentage of asteroid retrograde rotation & Varying from 50 to 100\% by 10\% increments \\  \hline
 Asteroid rotational period & Maxwellian 3-20 hr with peak at 6 hr \\
 \hline
 Only considered diurnal component of Yarkovsky Drift &  Dominates over seasonal component \\ \hline
Asteroid dynamical ellipticity $\Delta$ & Gaussian with $\mu = 0.25$ and $\sigma = 0.05$ \\ \hline
 Preference for YORP to accelerate or decelerate asteroid spin & 50:50 and 80:20 (acceleration:deceleration)  \\
 \hline
\end{tabular}
\caption{Note. The asteroid physical parameters are based on typical values for C-type asteroids \citep{2015aste.book..297N}.  The dynamical properties stem from current predictions of YORP theory (see \citeauthor{2002Icar..159..449V} \citeyear{2002Icar..159..449V}; \citeauthor{ 2004Icar..172..526C} \citeyear{2004Icar..172..526C}; \citeauthor{2005AJ....130.1267B} \citeyear{2005AJ....130.1267B}; \citeauthor{2006AREPS..34..157B} \citeyear{2006AREPS..34..157B}; \citeauthor{2015aste.book..509V} \citeyear{2015aste.book..509V}).} \label{fig:fig6}
\end{center}
\end{table}

\newpage

\section{Analysis} \label{sec:Analysis}

We simulated 500 test bodies over 150 Myr to model the past evolution of the synthetic Clarissa family.  For each body, we compute the synthetic 
proper elements with 0.5 Myr cadence and 10 Myr Fourier window 
\citep{SidlichovskyNesvorny1996}. Our goal is to match the orbital distribution of the real Clarissa family. This is done as
follows. The top panel of Figure \ref{fig:fig3} shows the semimajor axis distribution of 114 members of the Clarissa family with the sizes $D \simeq 2$ km. We denote the number of known asteroids in each of the bins as $N^{i}_{\mathrm{obs}}$, where $i$ spans 25 bins as shown in Figure \ref{fig:fig3}.  We use one million trials 
to randomly select 114 of our synthetic family asteroids and compute their semimajor axis distribution $N^{i}_{\mathrm{mod}}$ for the same bins.  For each of 
these trials we compute a $\chi^2$-like quantity, 
\begin{equation} \label{eq:eqchi}
    \chi^2_a(T) = \sum_{i}^{} \frac{\left(N^i_{\mathrm{mod}} - N^i_{\mathrm{obs}}\right)^2}{N^i_{\mathrm{obs}}},
\end{equation}
where the summation goes over all 25 bins.  The normalization factor of $\chi^2_a(T)$, namely $\sqrt{N^i_{\mathrm{obs}}}$, is a formal statistical uncertainty 
of the population in the $i$th bin.  We set the denominator equal to unity if $N^i_{\mathrm{obs}} = 0$ in a given bin.

\begin{figure}[ht!]
\epsscale{1.1}
\plotone{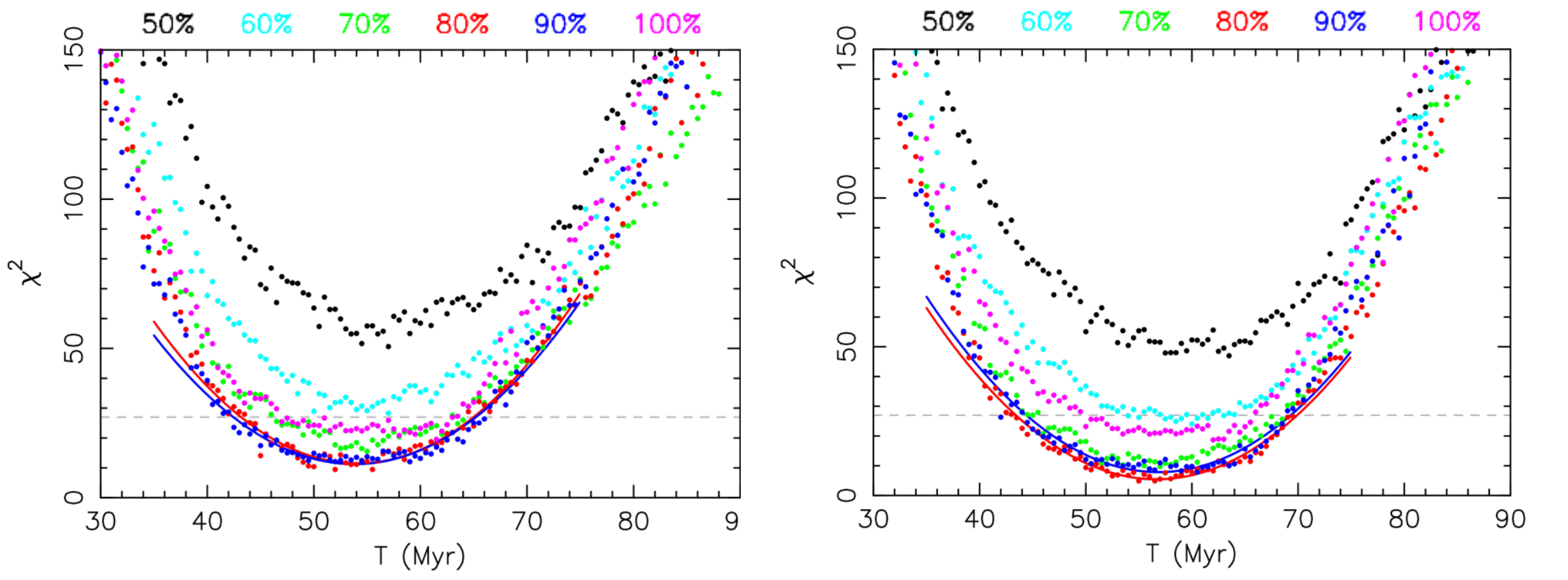}
\caption{Time-dependence of the $\chi^2$ function defined in Eq. (\ref{eq:chisqfull}): left panel for the YORP1 model, right panel for the YORP2 model.  These simulations used a 20 m s$^{-1}$ initial ejection velocity field that is isotropic in space.  Symbols show $\chi^2$ computed by 
the procedure described in Section \ref{sec:Analysis}.  The color coding corresponds to models with different fractions of initially retrograde rotators (see Figure \ref{fig:fig8} for a summary of the parameters): 50\% (black), 60\% (cyan), 70\% (green), 80\% (red), 90\% (blue), and 100\% (purple).  The red and blue parabolas are 
quadratic fits of $\chi^2$ near the minima of the respective data sets.  The dashed light-gray line marks the value of 27, equal to the number 
of effective bins.  The best age solutions are $54\pm 6$ Myr and $56 \pm 6$ Myr for YORP1 and YORP2, respectively. The YORP2 model provides the best solution with $\chi^2 \simeq 5.4$.  In both cases, simulations with 70\%-90\% of initially 
retrograde rotators provide the best solutions.}\label{fig:fig7}
\end{figure}

Another distinctive property of the Clarissa family is the distribution of eccentricities sunward of the J4-S2-1 resonance. 
We denote as $N^+_{\mathrm{obs}}$ the number of members with $a_{\rm p}<2.398$ au and $e_{\rm p}>0.1065$ (i.e., sunward from J4-S2-1 and eccentricities larger 
than the proper eccentricity of (302) Clarissa; Figure \ref{fig:fig2}). Similarly, we denote as $N^-_{\mathrm{obs}}$ the number of members with $a_{\rm p}<2.398$ au 
and $e_{\rm p}<0.1065$. For $D\simeq2$ km, we find $N^+_{\mathrm{obs}} = 48$ and $N^-_{\mathrm{obs}} = 5$. It is peculiar that $N^+_{\mathrm{obs}}/N^-_{\mathrm{obs}} \simeq 10$ 
because the initial family must have had a more even distribution of eccentricities. This has something to do with crossing of the J4-S2-1 
resonance (see below). As our goal is to simultaneously match the semimajor axis distribution and $N^+_{\mathrm{obs}}/N^-_{\mathrm{obs}}$, we define
\begin{equation} \label{eq:chisqfull}
    \chi^2(T) = \chi^2_a(T) + \frac{\left(N^+_{\mathrm{mod}}-N^+_{\mathrm{obs}} \right)^2}{N^+_{\mathrm{obs}}} + \frac{\left(N^-_{\mathrm{mod}}-N^-_{\mathrm{obs}}\right)^2}{N^-_{\mathrm{obs}}} \ ,
\end{equation}
where $N^-_{\mathrm{mod}}$ and $N^+_{\mathrm{mod}}$ are computed from the model.
The Clarissa family age is found by computing the minimum of $\chi^2(T)$ (Figure \ref{fig:fig7}).

We find that $\chi^2(T)$ always reaches a single minimum in $0<T<150$ Myr. The minimum of $\chi^2(T)$ was then determined by visual inspection, performing a second-order polynomial fit of the form $\chi^2(T) \simeq aT^2+bT+c$ in the vicinity of the minimum, and thus correcting the guessed value to $T_{*} = -b/2a$.  After inspecting the behavior of $\chi^2(T)$, we opted to use a $\pm$15 Myr interval where we fit the second-order polynomial; for instance, between 40 and 70 Myr, if the minimum is at 55 Myr, and so on.  A formal uncertainty is found by 
considering an increment $\Delta \chi^2$ resulting in $\delta T = \sqrt{\Delta \chi^2/a}$.  Thus the age of the Clarissa family is 
$t_{\rm age}=T_{*} \pm \delta T$. In a two-parameter model, where the two parameters are $T$ and the initial fraction of retrograde rotators, we need 
$\Delta \chi^2 = 2.3$ for a $68\%$ confidence limit or 1$\sigma$, and $\Delta \chi^2 = 4.61$ for a 90\% confidence limit or 2$\sigma$ (\citeauthor{2007Press} \citeyear{2007Press}, Chapter~15, Section~6). 
Our error estimates are approximate. The model has many additional parameters, such as the initial velocity field, thermal inertia, bulk density, etc.
Additionally, a Kolmogorov-Smirnov two-sample test (\citeauthor{2007Press} \citeyear{2007Press}, Chapter~14, Section~3) was performed on selected models (Figures \ref{fig:fig9} and \ref{fig:fig10}).  This test provides an alternative 
way of looking at the orbital distribution of Clarissa family members in the semimajor axis. It has the advantage of being independent of binning. 


\section{Results} \label{sec:results}

\subsection{Isotropic velocity field with 20 m s$^{-1}$} \label{sec:iso20}
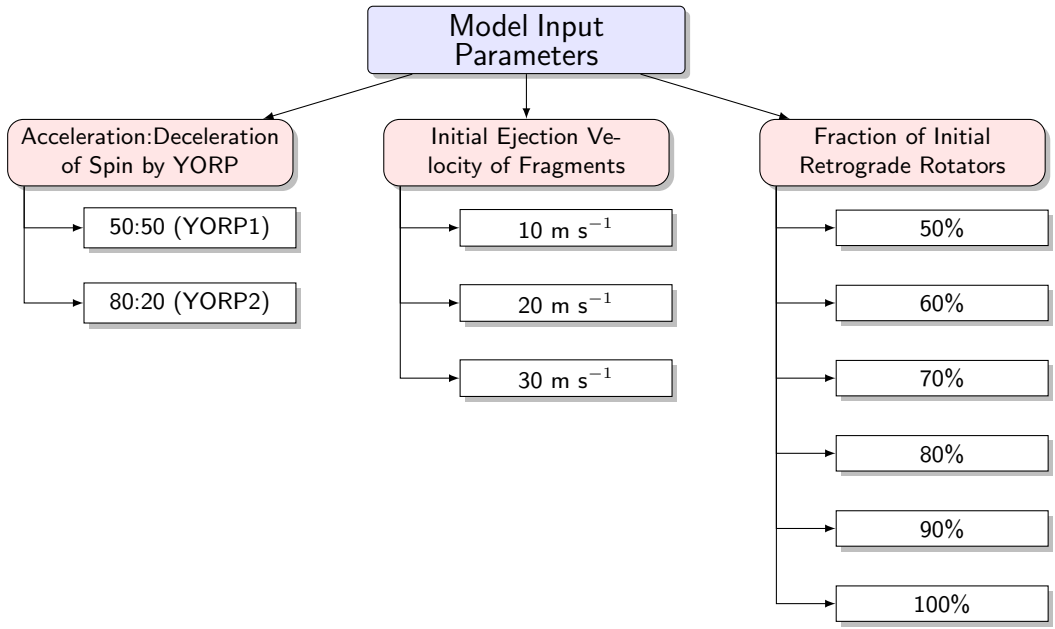
\begin{figure}[ht!]
\begin{center}
\begin{tikzpicture}[
  level 1/.style={sibling distance=50mm},
  edge from parent/.style={->,draw},
  >=latex]

\node[root] {\large Model Input Parameters}
  child {node[level 2] (c1) {Acceleration:Deceleration of Spin by YORP}}
  child {node[level 2] (c2) {Initial Ejection Velocity of Fragments}}
  child {node[level 2] (c3) {Fraction of Initial Retrograde Rotators}};

\begin{scope}[every node/.style={level 3}]
\node [below of = c1, xshift=15pt] (c11) {50:50 (YORP1)};
\node [below of = c11] (c12) {80:20 (YORP2)};

\node [below of = c2, xshift=15pt] (c21) {10 m s$^{-1}$};
\node [below of = c21] (c22) {20 m s$^{-1}$};
\node [below of = c22] (c23) {30 m s$^{-1}$};

\node [below of = c3, xshift=15pt] (c31) {50\%};
\node [below of = c31] (c32) {60\%};
\node [below of = c32] (c33) {70\%};
\node [below of = c33] (c34) {80\%};
\node [below of = c34] (c35) {90\%};
\node [below of = c35] (c36) {100\%};
\end{scope}

\foreach \value in {1,2}
  \draw[->] (c1.195) |- (c1\value.west);

\foreach \value in {1,2,3}
  \draw[->] (c2.195) |- (c2\value.west);

\foreach \value in {1,...,6}
  \draw[->] (c3.195) |- (c3\value.west);
\end{tikzpicture}
\end{center}
\caption{Summary of all input parameters for the YORP model simulations as shown in Figure \ref{fig:fig7} and explained in Section \ref{sec:results}. This diagram can be read as follows:  for example, in the YORP1 model (50:50 chance of acceleration:deceleration of spin by YORP) we simulated fragments with an initial ejection velocity of 20 m s$^{-1}$ which included varying fractions of initially retrograde rotators from 50\% to 100\% in increments of 10\%. The initial ejection velocity of 20 m s$^{-1}$ represents overall 12 simulations with the YORP1 and YORP2 models.  These simulated cases include only the isotropic velocity field.} \label{fig:fig8}
\end{figure}

These reference jobs used the assumption of an initially isotropic velocity field with all fragments launched at 20 m~s$^{-1}$ with respect to (302) 
Clarissa.  This set of simulations included two cases: the (i) YORP1 model (equal likelihood of 
acceleration and deceleration of spin), and (ii) the YORP2 model (80\% chance of acceleration versus 20\% chance of deceleration).  In each case, 
we simulated six different scenarios with different percentages of initially retrograde rotators (from 50\% to 100\% in increments of 
10\%).  See Figure \ref{fig:fig8} for a summary diagram of model input parameters.  In total, this effort represented 12 simulations, each following 500 test Clarissa family members.

Figure \ref{fig:fig7} summarizes the results by reporting the time dependence of $\chi^2(T)$ from Eq. (\ref{eq:chisqfull}). In all cases, $\chi^2(T)$ 
reaches a well defined minimum.  Initially the test body distribution is very different from the orbital structure of the Clarissa family
and $\chi^2(0)$ is therefore large.  For $T \geq$ 100 Myr, the simulated bodies evolve too far from the center of the family, well beyond 
the width of the Clarissa family in $a_{\rm p}$, and $\chi^2(T)$ is large again. The minimum of $\chi^2(T)$ occurs near 50-60 Myr. For the models with
equal split of prograde and retrograde rotators (Figure \ref{fig:fig7}, black symbols) the minimum $\chi^2(T_{*}) \simeq 50$, which is inadequately
large for 27 data points (this applies to both the YORP1 and YORP2 models).  This model can therefore be rejected. The main deficiency of this model
is that bodies have an equal probability to drift inward or outward in $a_{\rm P}$ (left panel of Figure \ref{fig:fig11}). The model therefore produces a symmetric distribution 
in semimajor axis, which is not observed (see the top panel of Figure \ref{fig:fig3}). The simulations also show that the M1/2 orbital resonance with 
Mars is not strong enough to produce the observed asymmetry. We thus conclude that the $a_{\rm p}$ distribution asymmetry must be a consequence of 
the predominance of retrograde rotators in the family. This prediction can be tested observationally. 

The results shown in Figure \ref{fig:fig7} indicate that the best solutions are obtained when 70\%-90\% of fragments have initially retrograde
rotation. These models lead to $\chi^2(T_{*}) \simeq 11.4$ for YORP1 and $\chi^2(T_{*}) \simeq 5.4$ for YORP2.  Both these values are acceptably low. A statistical test shows that the probability $\chi^2$ should attain or exceed this level by random fluctuations is greater than 90\% (\citeauthor{2007Press} \citeyear{2007Press}, Chapter~15, Section~2).  The inferred
age of the Clarissa family is $t_{\rm age}=54 \pm 6$ Myr for the YORP1 model and $t_{\rm age}=56 \pm 6$ Myr for the YORP2 model. 

\begin{figure}[ht!]
\epsscale{0.7}
\plotone{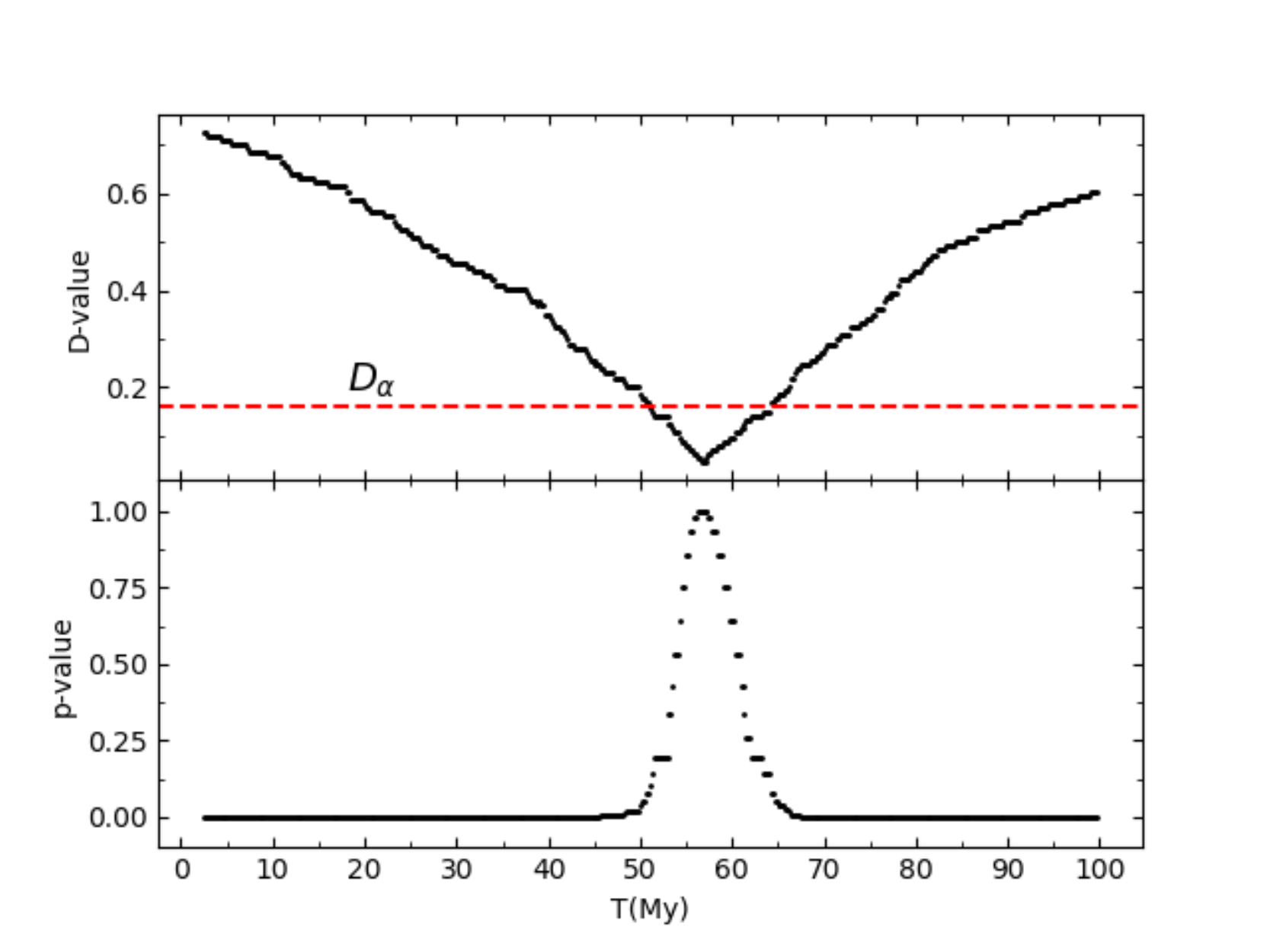}
\caption{Kolmogorov-Smirnov two-sample test: $D$-value vs. time (top panel), and $p$-value vs. time (bottom panel), both computed from the cumulative 
distribution functions $F_1(a_{\rm p})$ (model) and $F_2(a_{\rm p})$ (observed).  The test was applied to the preferred YORP2 model (80\% preference for acceleration 
by YORP with 80\% of retrograde rotators; Fig. \ref{fig:fig7}). The dashed red line refers to the critical $D$-value, $D_{\alpha}$ ($\alpha = 0.10$), which corresponds to a 90\% 
confidence limit (\citeauthor{PracticalEngineering} \citeyear{PracticalEngineering}, appendix 3). We find $t_{\rm age}=56^{+7}_{-6}$ Myr using this test. This result closely matches that obtained 
with the $\chi^2$ method described in Section \ref{sec:Analysis}.} \label{fig:fig9}
\end{figure}

\begin{figure}[ht!]
\epsscale{0.6}
\plotone{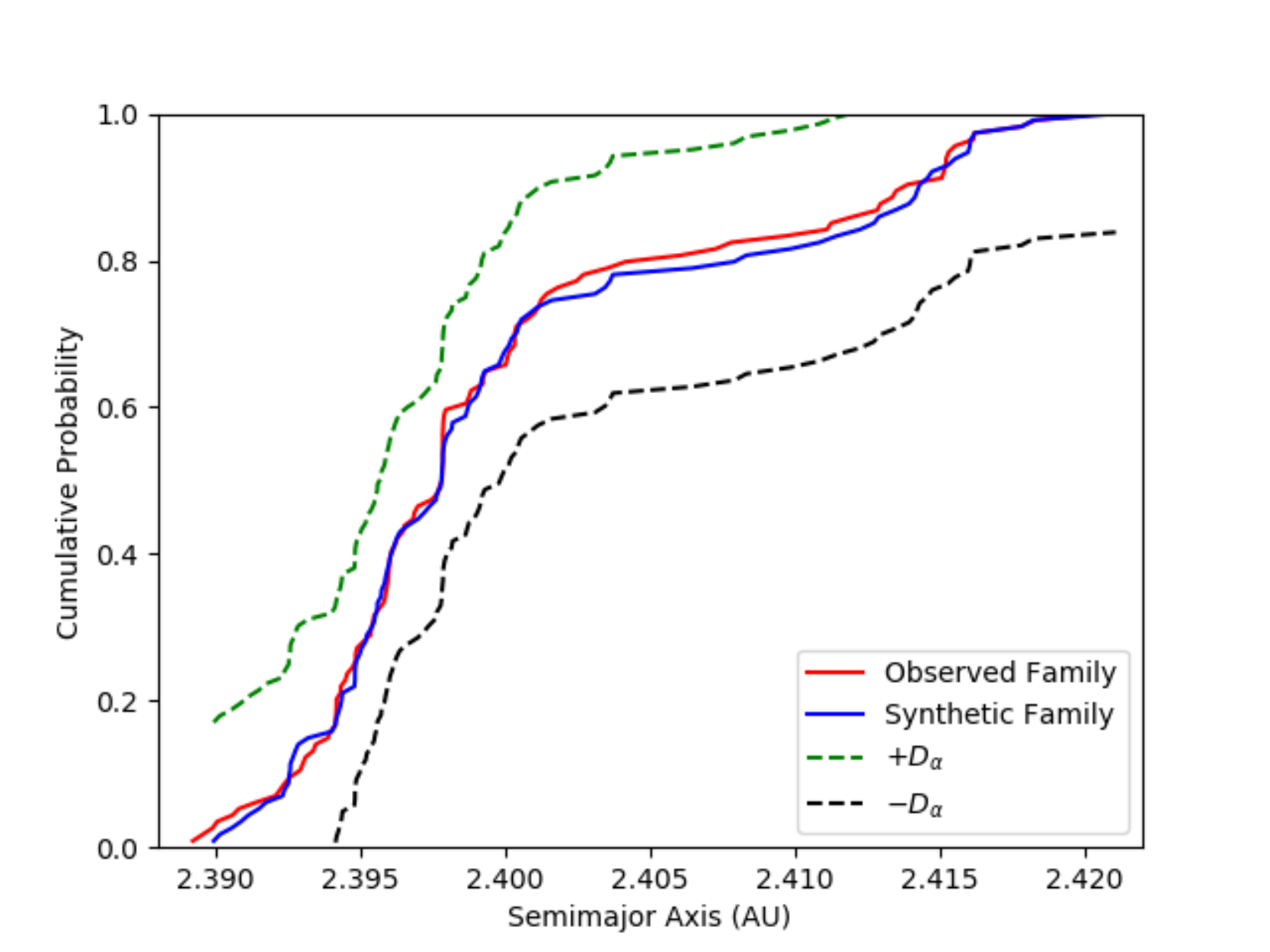}
\caption{Kolmogorov-Smirnov two-sample test: cumulative distribution functions $F_1(a_{\rm p})$ (model, blue line) and $F_2(a_{\rm p})$ (observed, red line). The model distribution is shown for our 
preferred YORP2 model at $T_* = 56$ Myr. This curve (blue) corresponds to the minimum of the $D$-value and the maximum $p$-value plotted in Figure \ref{fig:fig9}.  
The green and black dashed lines are $F_1(a_{\rm p})$ $\pm$ $D_{\alpha}$ corresponding a 90\% confidence band where $F_1(a_{\rm p})-D_{\alpha} \leq F_2(a_{\rm p}) \leq F_1(a_{\rm p}) + D_{\alpha}$.  This interval contains the true cumulative distribution function $F_2(a_{\rm p})$ with a 90\% probability. }\label{fig:fig10}
\end{figure}

\begin{figure}[ht!]
\epsscale{1.1}
\plotone{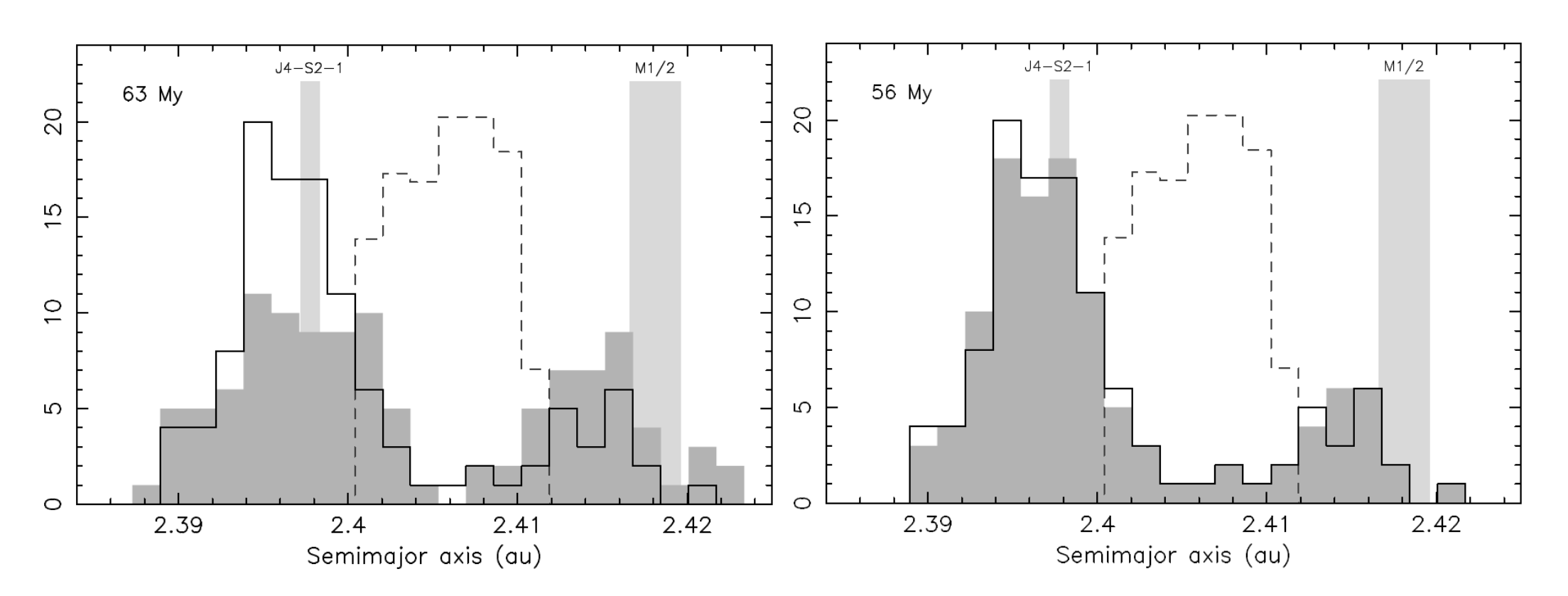}
\caption{Minimum $\chi^2$ solution for the  YORP2 model in both panels with 20 m s$^{-1}$ ejection speed with 50\% of initially retrograde spins (left panel) and 80\% of initially 
retrograde spins (right panel). The solid black line represents the observed family corresponding to the top panel of Figure \ref{fig:fig3}.  The dashed line 
is the initial distribution of test bodies.  The gray histogram is the model distribution where $\chi^2$ reached a minimum in the simulation corresponding to age solutions of $T_{*}$ = 63~Myr (left panel) and 56~Myr (right panel).  In the left panel the model distribution is quite symmetric and does not match the observed family distribution.  There is only a slight asymmetry in the model distribution (left panel) which is due to asteroids leaking out of the family range via the M1/2 resonance.  The light-gray bars highlight locations of the principal resonances.}
\label{fig:fig11}
\end{figure}

The results of the Kolmogorov-Smirnov test confirm these inferences. For example, if we select a 90\% confidence limit to be able to compare with the best-fit $\chi^2$ 
result, we obtain $t_{\rm age}= 56^{+7}_{-6}$ Myr for the YORP2 model (see Figs. \ref{fig:fig9} and \ref{fig:fig10}). The best fit to the 
observed $a_{\rm p}$ distribution is shown for YORP2 in Figure \ref{fig:fig11} (right panel). The model distribution for $T_{*} = 56$ Myr indeed represents an excellent 
match to the present Clarissa family. 

\begin{figure}[ht!]
\epsscale{0.6}
\plotone{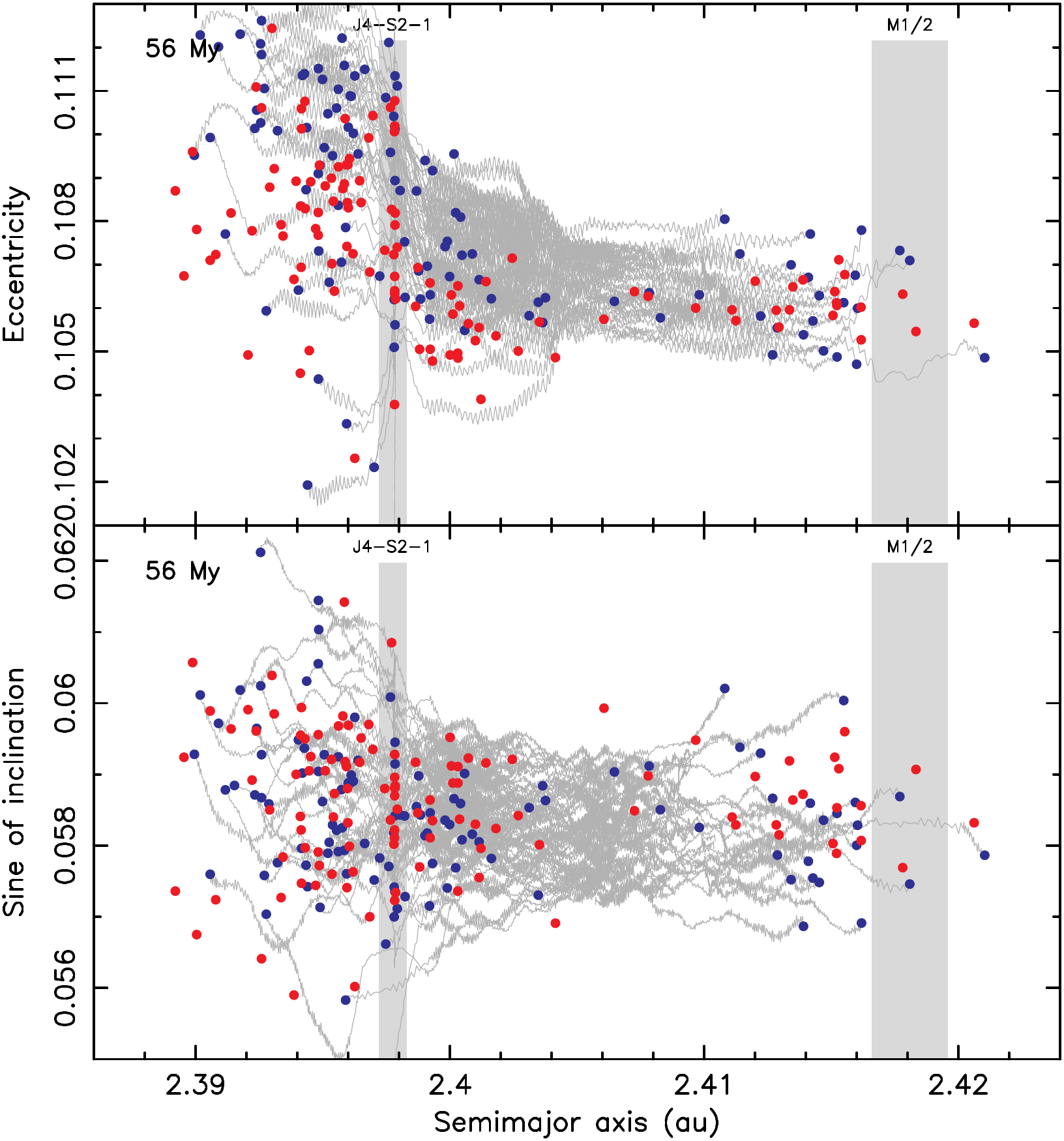}
\caption{Orbital evolution of family members in our preferred YORP2 model 
shown in Fig. \ref{fig:fig11} (right panel). The proper orbital elements $e_{\rm p}$ and $a_{\rm p}$ are shown in the top panel, and 
$\sin{i_{\rm p}}$ and $a_{\rm p}$ are shown in the bottom panel.  The red symbols are the 114 members of the Clarissa family with sizes $D \simeq 2$ km (those within 
the strip delimited by the dashed gray lines in the bottom panel of Figure \ref{fig:fig3}). The blue symbols show orbits of 114 modeled bodies for $T_{*} = 56$ Myr.  The gray lines show the evolutionary tracks of the test bodies.}\label{fig:fig12}
\end{figure}

The orbital distribution produced by our preferred YORP2 model is compared with observations in Figure \ref{fig:fig12}. We note that the test bodies
crossing the J4-S2-1 resonance often have their orbital eccentricity increased. This leads to the predominance of orbits with $e_{\rm p}>0.1065$ 
for $a_{\rm p}<2.398$ au.  We obtain $N^+_{\mathrm{mod}} = 45$ and $N^-_{\mathrm{mod}} = 5$, which is nearly identical to the values found in the real family 
($N^+_{\mathrm{obs}} = 48$ and $N^-_{\mathrm{obs}} = 5$). Suggestively, even the observed $\sin{i_{\rm p}}$ distribution below the J4-S2-1 resonance, which is slightly wider, is well reproduced. We also note a hint of a very weak mean motion resonance at $a_{\rm P}\simeq 2.404$~au which manifests itself as a slight dispersal of $e_{\rm P}$. Using tools discussed and provided by \cite{2014Icar..231..273G}, we tentatively identified it as a three-body resonance J6-S7-1, but we did not prove it by analysis of the associated critical angle.

\subsection{Anisotropic velocity field with 20 m s$^{-1}$}

Here we discuss (and rule out) the possibility that the observed asymmetry in $a_{\rm p}$ is related to an anisotropic ejection velocity field (rather than to 
the preference for retrograde rotation as discussed above). To approximately implement an anisotropic velocity field we select test bodies 
initially populating the left half of the green ellipses in Fig. \ref{fig:fig2} (i.e., all fragments assumed to have initial $a_{\rm p}$ lower than 
(302) Clarissa) and adopt a 50-50\% split of prograde/retrograde rotators. This model does not work because the evolved distribution in 
$a_{\rm p}$ becomes roughly symmetrical (with only a small sunward shift of the center). This happens because the effects of the Yarkovsky drift  
on the $a_{\rm p}$ distribution are more important than the initial distribution of fragments in $a_{\rm p}$.

We also tested a model that combined the preference for retrograde rotation with the anisotropic ejection field. As before,
we found that the best-fitting models were obtained if there was an $\sim$80\% preference for retrograde rotation. The fits were not 
as good, however, as those obtained for the isotropic ejection field. The minimum $\chi^2(T_{*})$ achieved was $\simeq$ 12, which is significantly
higher than the previous result with $\chi^2(T_{*}) \simeq 5.4$.  We therefore conclude that the observed structure of the Clarissa family 
can best be explained if fragments were ejected isotropically and there was $\simeq$4:1 preference for retrograde rotation. This represents an important constraint on the impact that produced the Clarissa family and, more generally, on the physics of large-scale collisions. 


\subsection{Isotropic velocity field with 10 and 30 m s$^{-1}$}

\begin{figure}[ht!]
\epsscale{1.1}
\plotone{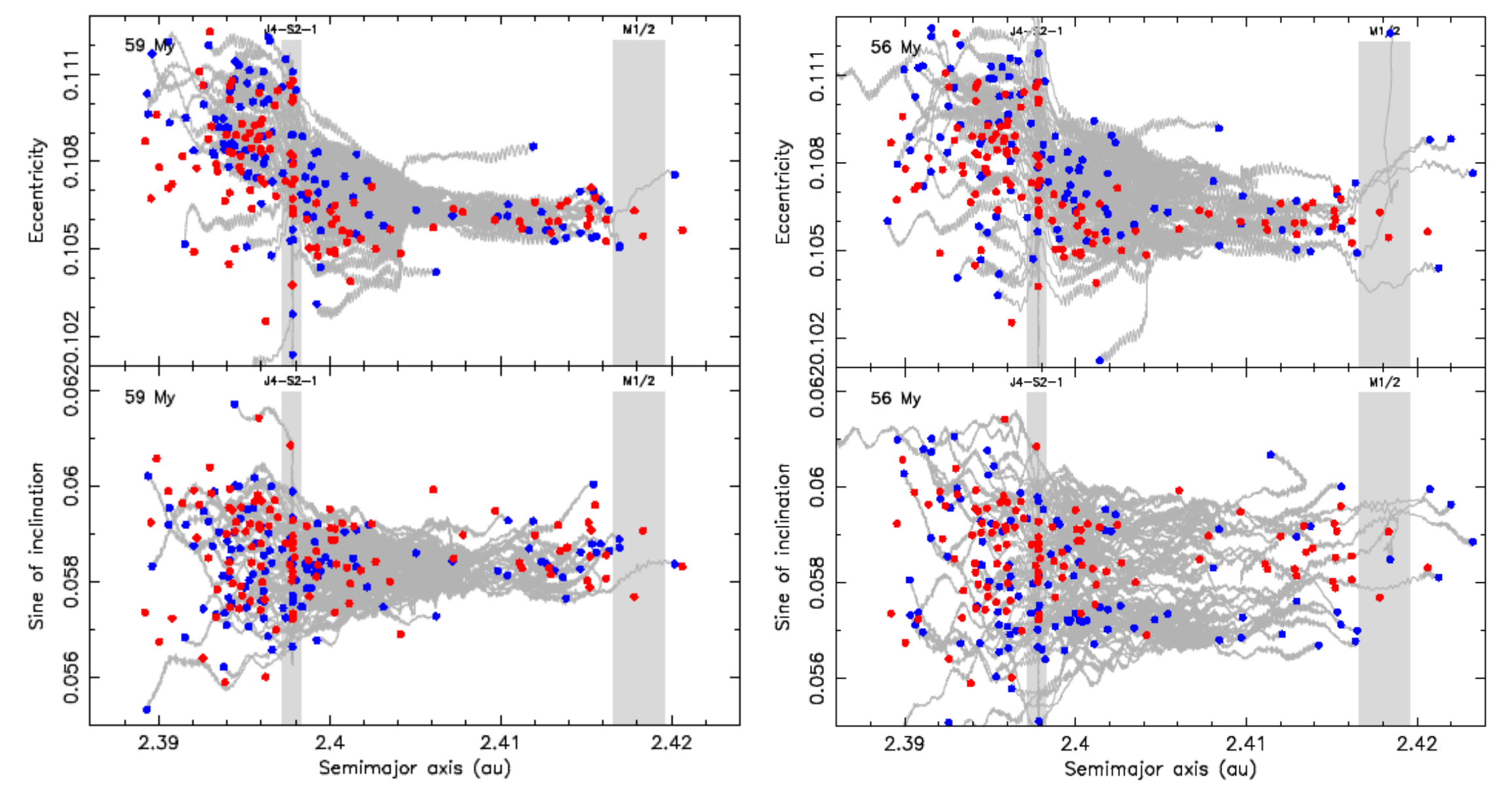}
\caption{Orbital evolution of family members in our preferred YORP2 model (80\% of initially retrograde spins and 80\% preference for spin acceleration by YORP) for initial ejection velocities of $v_{\rm ej}$ = 10 m s$^{-1}$(left panel) and $v_{\rm ej}$ = 30 m s$^{-1}$ (right panel).  This figure is similar to Figure \ref{fig:fig12} but with two different ejection speeds. The red symbols are the 114 members of the Clarissa family with sizes $D \simeq 2$ km (those within 
the strip delimited by the dashed gray lines in the bottom panel of Figure \ref{fig:fig3}). The blue symbols show orbits of 114 modeled bodies at the time of minimum $\chi^2$ of the respective simulation, $T_{*} = 59$ Myr (left panel) and $T_{*} = 56$ Myr (right panel).  The gray lines show the evolutionary tracks of the test bodies in both simulations.}\label{fig:13}
\end{figure}

We performed additional simulations with the isotropic ejection field and velocities of 10 and 30 m s$^{-1}$. The main goal of these simulations
was to determine the sensitivity of the results to this parameter. Analysis of the simulations with 10 m s$^{-1}$ revealed results 
similar to those obtained with 20 m s$^{-1}$ (Section \ref{sec:iso20}). For example, the best-fitting solution for the preferred YORP2 
had $\chi^2(T_{*}) \simeq 3.9$ and $t_{\rm age}=59 \pm 5$~Myr (90\% confidence interval). Again,  70\% to 90\% of test bodies are required to have initially retrograde rotation.  Results obtained with 30 m s$^{-1}$ showed that this value is already too large to provide an acceptable fit.  The best-fit solution of all investigated models with 30 m s$^{-1}$ is $\chi^2 \simeq 25$. This happens because the initial spread in the semimajor axis is too large and the Yarkovsky and YORP effects are not capable of producing Clarissa family ears (see Fig. \ref{fig:fig3}). We conclude
that ejection speeds $\gtrsim 30$  m s$^{-1}$ can be ruled out.  Figure \ref{fig:13} shows results similar to Figure \ref{fig:fig12}, but for ejection velocities $v_{\rm ej}=10$ m~s$^{-1}$ (left
panel) and $v_{\rm ej}=30$ m~s$^{-1}$ (right panel); all other simulation parameters are the same as in Fig. \ref{fig:fig12} and the configuration
is shown when $\chi^2$ of the corresponding run reached a minimum.  The former simulation, $10$ m~s$^{-1}$ ejection speed, still provides very good results. The initial spread in proper eccentricities near (302)~Clarissa (at $a_{\rm P}\simeq 2.4057$~au) is significantly smaller, but the above-mentioned weak mean motion resonance at $2.404$~au suitably extends the family at smaller $a_{\rm P}$
region as the members drift across. This helps to balance the $e_{\rm P}$ distribution of the family members below the J4-S2-1 resonance and also provides a tight $e_{\rm P}$ distribution near the M1/2 resonance. On the other hand, the simulation with a $30$ m~s$^{-1}$ ejection speed gives much worse results (the best we could get was $\chi^2(T_\star)\simeq 23.8$ for the simulation shown on the right panel of Fig.\ref{fig:13}). Here the family initial extension in $e_{\rm P}$ and $\sin i_{\rm P}$ is large, and this implies that also the population of fragments that crossed the J4-S2-1 resonance remains unsuitably large and contradicts the evidence of a shift toward larger values on its sunward side.  

\section{Discussion and Conclusions}

The Clarissa family is an interesting case.  The family's location in a dynamically quiet orbital region of the inner belt allowed us to model 
its structure in detail. Its estimated age is older than any of the very young families (e.g., 
\citeauthor{2002Natur.417..720N} \citeyear{2002Natur.417..720N, 2006IAUS..229..289N, 2006Sci...312.1490N}) but younger than any of the families to which the Yarkovsky effect chronology
was previously applied (e.g., \citeauthor{2006Icar..182..118V}  \citeyear{2006Icar..182..118V,2006Icar..183..349V, 2006Icar..182...92V}). Specifically, we found that the Clarissa family is $56 \pm 6$~Myr 
old (formal 90\% confidence limit).  The dependence on parameters not tested in this work may imply a larger uncertainty. For example, here
we adopted a bulk density $\rho=1.5$ g cm$^{-3}$. In the case of pure Yarkovsky drift the age scales with $\rho$ as $t_{\rm age}\propto\rho$; higher/lower densities would thus imply 
older/younger ages.  However, in our model this scaling is more complicated since altering $\rho$ changes the YORP timescale and the speed of resonance crossing. 

The initial ejection velocities were constrained to be smaller than $\simeq$ 20 m s$^{-1}$, a value comparable to the escape velocity from (302) Clarissa.  
We found systematically better results for the model where $\sim$80\% of fragments had rotation accelerated by YORP and the remaining $\sim$20\% 
had rotation decelerated by YORP. This tendency is consistent with theoretical models of YORP and actual YORP detection, which suggest the same 
preference (as reviewed in \citeauthor{2015aste.book..509V} \citeyear{2015aste.book..509V}).  

The most interesting result of this work is the need for asymmetry in the initial rotation direction for small fragments.  
We estimate that between 70\% and 90\% of $D \simeq$ 2 km Clarissa family members initially had retrograde rotation. As this preference was not 
modified much by YORP over the age of the Clarissa family, we expect that the great majority of small family members with $a<2.406$ au (i.e., lower 
than the semimajor axis of (302) Clarissa) must be retrograde rotators today. This prediction can be tested observationally.

In fact, prior to running the test cases mentioned in Figure \ref{fig:fig8} of Section \ref{sec:results} we expected that simulating more retrograde rotators in roughly the 80:20 proportion would match the distribution of the observed Clarissa family.  We see roughly the same proportion in the V-shape of Figure \ref{fig:fig3} where there are more asteroids on the left side. Possible causes for the split of prograde/retrograde rotators in the Clarissa family and other asteroid families could be a consequence of the original parent body rotation, the geometry of impact, 
fragment reaccumulation, or something else.
 Some previously studied asteroid families have already hinted at possible asymmetries or peculiar diversity.  For example, the 
largest member of the Karin family is a slow prograde rotator, while a number of members following (832) Karin in size are retrograde rotators (\citeauthor{2004Icar..170..324N} \citeyear{2004Icar..170..324N}; \citeauthor{2016AJ....151..164C} \citeyear{2016AJ....151..164C}).  Similarly, the largest member of the Datura family is a very fast prograde rotator, while several members with smaller size are very slowly rotating and peculiarly shaped objects, all in a prograde sense (e.g., \citeauthor{2017A&A...598A..91V} \citeyear{2017A&A...598A..91V}).
The small members of the Agnia family are predominantly retrograde ($\simeq$ 60\%; \citeauthor{2006Icar..183..349V} \citeyear{2006Icar..183..349V}).  The inferred conditions of Clarissa family formation, together with (302) Clarissa's slow and retrograde rotation, therefore present an additional interesting challenge
for modeling large-scale asteroid impacts. We encourage our fellow researchers to investigate the interesting scientific problem of the possible causes in the split of prograde/retrograde rotators.

\appendix 
\section{Choice of parameters $\omega$ and $\lowercase{f}$} \label{Ap:A}
Obviously our choice of $\omega$ and $f$ for the initial configuration of the synthetic Clarissa family is not unique.
However, we argue that (i) there are some limits to be satisfied, and (ii) beyond these limits the results would
not critically depend on the choice of $f$ and $\omega$. First, we postulate an initial ejection velocity of
family members around $20$ m~s$^{-1}$ (about the escape velocity from (302) Clarissa) as the most probable value
(often seen in young asteroid families). Then, the choice $\omega + f$ either near 0$^{\circ}$ or 180$^{\circ}$
is dictated by the Clarissa family extent in proper inclination values in between the J4-S2-1 and M1/2 resonances
(see the green ellipses in Fig. \ref{fig:fig2}). There are no dynamical effects in between these two resonances
to increase the inclination to observed values. So, for instance, if $\omega + f$ were close to 90$^\circ$, the
spread in proper inclination would collapse to zero which is contrary to observation (see the
corresponding Gauss equation from \citeauthor{1996Icar..124..156Z} \citeyear{1996Icar..124..156Z}). In the same way,
if we want to have fragments roughly equally represented around Clarissa in the ($a_{\rm p}$, $e_{\rm p}$) plot (Fig. \ref{fig:fig2})
then we need $f$ near 90$^\circ$ (for instance, values of 0$^{\circ}$ or 180$^\circ$ would shrink the appropriate ellipse
to a line segment, again not seen in the data).

\section{Details of code extension} \label{Ap:B}
Here we provide few more details about implementation of radiation torques (the YORP effect) in our code; more can
be found in \cite{2017AJ....153..172V} and \cite{2015Icar..247..191B}. We do not assume a constant rate in rotational
frequency $\omega$, and we do not assume a constant obliquity $\epsilon$ of all evolving Clarissa fragments.
Instead, after setting some initial values ($\omega_0$, $\epsilon_0$) for each of them, we numerically integrate Eqs.~(3)
and (5) of \cite{2004Icar..172..526C}, or also Eqs.~(2) and (3) of Appendix~A in \cite{2015Icar..247..191B}. This means
$\frac{d\omega}{dt} = f(\epsilon)$ and $\frac{d\epsilon}{dt} = g(\epsilon)/\omega$. As a template for the $f$- and $g$-functions
we implement results from Figure~8 of \cite{2004Icar..172..526C} (note this also fixes the general dependence of the $g$-functions
on the surface thermal conductivity). The $g$-function has a typical wave pattern making obliquity evolve asymptotically
to 0$^{\circ}$ or 180$^\circ$ from a generic initial value. The $f$-function also has a wave pattern, though the zero value
is near $\sim 55^{\circ}$ and $\sim 125^{\circ}$ obliquity and at 0$^{\circ}$ or 180$^\circ$ obliquity \citep{2004Icar..172..526C}. \citeauthor{2004Icar..172..526C} had an equal likelihood of acceleration or deceleration of $\omega$ (due to the simplicity of their approach). This is our YORP1 model. When we tilt these statistics to 80\% asymptotically accelerating and 20\% asymptotically decelerating cases for $\omega$, we obtain our YORP2 model \citep[this is our empirical implementation of the physical effects studied in][]{2012ApJ...752L..11G}.
We can do that straightforwardly because at the beginning of the integration we assign to each of the fragments one particular realization of the $f$- and $g$-functions from the pool covered by \cite{2004Icar..172..526C} results (their Figure~8). 

The behavior at the boundaries of the rotation rate $\omega$ also needs to be implemented: (i) the shortest allowed rotation period is set to $2$~hr (before fission would occur), and (ii) the longest is set at a $1000$~hr rotation period. Modeling of the rotation evolution at these limits is particularly critical for old families, such as Flora or even Eulalia (discussed in \citeauthor{2017AJ....153..172V} \citeyear{2017AJ....153..172V} and \citeauthor{2015Icar..247..191B} \citeyear{2015Icar..247..191B}), because small fragments reach the limiting values over a timescale much smaller than the age of the family. Fortunately this problem is not an issue for the Clarissa family due to its young age. Many fragments in our simulation just make it to these asymptotic limits. Note also that while slowing down $\omega$ by YORP, asteroids do not cease rotation and start tumbling (not implemented in detail in our code). For that reason, while arbitrary and simplistic, the $1000$~hr smallest-$\omega$ limit is acceptable. Our code inherits from \cite{2015Icar..247..191B} some approximate recipes on what to do in these limits. For instance, when stalled at an $1000$~hr rotation period (``tumbling phase''), the bodies are sooner or later assumed to receive a sub-catastrophic impact that resets their rotation state to new initial values. When the rotation period reaches $2$~hr, we assume a small fragment gets ejected by fission and the rotation rate resets to a smaller value. In both cases, an entirely new set of values for the $f$- and $g$-functions is chosen.

\acknowledgments

Support for this work is given by SSERVI-CLASS grant NNA14AB05A and NASA grant NNX17AG92G (V.L. and H.C.).  Further support was given by the NASA Florida Space Grant Consortium Fellowship (V.L.).  Coauthor support for this study was given by the NASA SSW program (D.N.) and by the Czech Science Foundation (grant 18-06083S) (D.V.).  

\bibliography{clarissa_DN}{}
\bibliographystyle{aasjournal}

\end{document}